\documentclass{aa}          
\usepackage{graphicx}
\begin{document}

   \title{Near Infrared Intraday Variability of Mrk 421}


   \author {A. C. Gupta$^1$, D. P. K. Banerjee$^2$, N. M. Ashok$^2$ \& U. C. Joshi$^2$}

   \offprints{A. C. Gupta}
   
   \institute 
{$^1$ Harish-Chandra Research Institute, Chhatnag Road, Jhunsi, Allahabad - 211 019, India \\
$^2$    Astronomy and Astrophysics Division, Physical Research Laboratory, Navrangpura, 
Ahmedabad - 380 009, India \\
email: agupta@mri.ernet.in, orion@prl.ernet.in, ashok@prl.ernet.in, joshi@prl.ernet.in}

\date{Received; Accepted}

\titlerunning{Variability of Mrk 421}
   

\abstract{We report results from  our monitoring of the BL Lac object Mrk 421 in the near-IR $J$ 
band. The observations, aimed at studying the intraday variability (IDV) of the object, 
were carried out systematically over an extended (and near-continuous) period of eight 
nights from the 1.2m  Mount Abu Infrared Telescope, India. There are limited studies 
for Mrk 421 in the $J$ band for  such an extended period. The observation epoch for this 
study (25 February - 5 March 2003) was chosen to significantly overlap other concurrent 
studies of Mrk 421 in the X-ray/$\gamma$-ray regions being conducted using the Rossi X-ray timing
explorer (RXTE) and the solar tower atmospheric Cherenkov effect experiment (STACEE). 
Hence these results could be useful for a multi-wavelength analysis of the variability 
behavior of Mrk 421. We find that Mrk 421 was quite active during the observed period and 
showed significant IDV and short term variability. A maximum variation of 0.89 magnitudes 
is seen over the entirety of the observed period. Flaring activity, with typical brightness 
variations of $\sim 0.4$, are also seen on several occasions. The extent of the variability 
observed by us is compared with the results of other similar studies of Mrk 421 in the $J$ 
band. 
}

\maketitle

\section{Introduction}

In this work we report our studies of the BL Lac object Mrk 421. BL Lac objects are a class 
of radio-loud active galactic nuclei (AGNs) and a subclass of blazars. They often show large 
and violent variations in the complete electromagnetic spectrum and their emission is strongly 
polarized. Their radiation at all wavelengths is predominantly nonthermal. Compared to radio 
frequencies, more rapid changes are seen to occur at optical and near-IR bands (Stein et al. 1976). 
Using an orientation-based unification scheme for radio-loud AGNs, a BL Lac object is considered 
to manifest itself when the relativistic jets emerging from the very core of the galaxy are 
pointing nearly towards the observer (Urry \& Padovani 1995). 

In general the spectral energy distribution (SED) of blazars  has two components. The first 
component peaks in the IR-to-optical region for the so called red blazars (or low energy 
blazars - LBL) while it peaks in the  UV/X-ray region for the blue blazars (or high energy 
blazars - HBL) (Padovani \& Giommi 1995).  The origin of the first component is attributed to 
synchrotron emission from high energy electrons in the relativistic jet. The second component, 
which extends up to $\gamma$-rays, peaks at GeV energies in LBLs and at TeV energies in HBLs. 
The high energy peak is generally attributed to Inverse Compton (IC) scattering of soft photons. 
However, the origin of the soft photons that seed the IC component of blazar spectrum is not a 
well understood  aspect in the study of blazars.

Mrk 421 was first noted to be an object with a blue excess which turned out to be an elliptical 
galaxy with a bright, point like nucleus (Ulrich et al. 1975). The object showed optical 
polarization and the  spectrum of the nucleus was seen to be featureless - an aspect common to 
BL Lac objects. It is a nearby BL Lac object (z = 0.031) which is  classified as an HBL source 
because the energy of the synchrotron peak in its SED is higher than 0.1 keV. It is the brightest 
BL Lac object at X-ray and UV wavelengths. Mrk 421 was the first extragalactic source discovered 
at TeV energies at the 6 $\sigma$ level by the Whipple group (Punch et al. 1992) which was 
confirmed by the high energy gamma ray astronomy (HEGRA) group (Petry et al. 1996). It is the 
first AGN detected by STACEE in the 140 GeV band (Boone et al. 2002). It is one of the AGNs 
detected by the energetic gamma ray experiment telescope (EGRET) instrument in the 30 MeV - 30 GeV 
energy range by the Compton gamma ray observatory (CGRO) (Thompson et al. 1995). The imaging Compton 
telescope (COMPTEL) has also detected Mrk 421 in the 10 - 30 MeV range at the 3.2 $\sigma$ level 
(Collmar et al. 1999).

Blazar variability can be broadly divided into 3 classes viz.  microvariability or intra-day 
variability (IDV), short term outbursts and long term trends. Significant variations in flux of 
a few tenths of a magnitude over the course of  a day or less is often known as microvariability. 
Short term outbursts can range from weeks to months and long term trends can have time scales of 
several years. The first convincing evidence of optical IDV was found in the  blazar BL Lacertae 
(Miller et al. 1989). Specifically, Mrk 421 has been studied for variability in all regimes of 
the electromagnetic spectrum. Large and fast variations are often found in the optical - e.g 
a large optical variation of 4.6 mag has been seen in Mrk 421 (Stein  et al., 1976) and rapid 
optical variability is well exemplified by the detection of a 1.4 magnitude brightness change in 
the object in a 2.5 hour period (Xie et al., 1988). 
In the near-IR -the region with which we are concerned here- there have  also been efforts to 
monitor variability in Mrk 421. An exhaustive compilation summarizing all the near-IR results 
of Mrk 421 stretching over three decades is given in the recent work by Fan \& Lin (1999). 
From this work, and also subsequent reports, we note that  there have been very few studies 
of micro and short-term variability of Mrk 421 in the $J$ band. The notable studies in this 
respect are those by Takalo et al. (1992), Makino et al. (1987) and Kidger et al. (1999). The 
study by Takalo et al. (1992) gives short-term variability in 2 nearby slots ( JD 2448272-2448280 
and 2448324-2448335) with  durations of 9 days and 12 days respectively. However their data gives 
one $J$ magnitude per night and the sampling is generally over alternate nights - a total of 12 $J$ 
magnitudes are reported. Makino et al's (1987) results constitute a part of an extensive 
multi-wavelength campaign. However we concern ourselves here with only their  $J$ band results 
which give single point data per night for  5 continuous nights. Kidger et al.(1999) have studied 
IDV over a period of 3 hours in one night. Thus there is limited work which studies both IDV and 
short term variability in the near-IR bands with a good sampling rate. In view of this we decided 
to pursue the present study which addresses both the IDV and short term variability of Mrk 421. 
In addition to the above motivation, we were also made aware (Bhat, 2003) that a multi-wavelength 
campaign using RXTE and STACEE was to be conducted for Mrk 421. We therefore have synchronized most 
of our observations with as many observation slots of the RXTE schedule as was possible. Hence the 
present $J$ band data give simultaneous information in an additional spectral window. These 
observations should therefore make it possible to search for correlations and time delays between 
different spectral regions. Thus the present work could give useful input to multi-wavelength 
modeling of the Blazar phenomenon aimed at understanding the causes for their variability.
     
\section{Observations and Data Reduction}

\vskip 2mm
\begin{table*}[h]
\caption[]{The log of observations for Mrk 421 and its derived magnitudes.
Some of the abbreviated titles are:  Exposure time represented by Exp. time,
 Total integration time by Int. time and  $J$ band magnitude by $J$ Mag.}

\begin{tabular}{cccccccccccc}
\hline \hline \\
JD  &  Exp.    & Int. & Airmass &  J Mag. &&&  JD  &  Exp.    & Int. & Airmass &  J Mag. \\
(2450000+) & Time(s)&time(s)& & (Error) && & (2450000+)&Time(s)&time(s)& & (Error)   \\
\hline
  {\bf Feb. 25} &&&&&&& 2701.378& 60 & 300 & 1.087 &   11.54(0.07) \\
  2696.393& 20 & 180 & 1.093 &	 11.20(0.03)&&&  2701.383& 60 & 300 & 1.097 &  11.45(0.07) \\ 
  2696.396& 20 & 180 & 1.099 &   11.18(0.03)&&&  {\bf Mar. 03} &&&& \\
  2696.399& 20 & 180 & 1.106 &   11.12(0.03)&&&  2702.302& 60 & 300 & 1.031 &  11.36(0.05) \\
  2696.402& 20 & 180 & 1.113 &   11.27(0.03)&&&  2702.306& 60 & 300 & 1.038 &  11.66(0.05) \\
  {\bf Feb. 26} &&&&&&&                          2702.312& 60 & 300 & 1.029 &  11.23(0.05) \\
  2697.281& 60 & 540 & 1.065 &   11.58(0.05)&&&  2702.317& 60 & 300 & 1.029 &  11.34(0.05) \\
  2697.287& 60 & 240 & 1.055 &   11.59(0.05)&&&  2702.321& 60 & 300 & 1.030 &  11.24(0.05) \\
  2697.292& 60 & 240 & 1.047 &   11.57(0.05)&&&  2702.327& 60 & 300 & 1.031 &  11.36(0.05) \\
  2697.298& 60 & 240 & 1.042 &   11.46(0.05)&&&  2702.333& 60 & 300 & 1.034 &  11.46(0.05) \\
  2697.331& 60 & 240 & 1.029 &   11.54(0.05)&&&  2702.341& 60 & 300 & 1.040 &  11.19(0.05) \\
  2697.335& 60 & 240 & 1.030 &   11.64(0.05)&&&  2702.357& 60 & 300 & 1.057 &  11.51(0.08) \\
  2697.339& 60 & 240 & 1.031 &   11.69(0.05)&&&  2702.362& 60 & 300 & 1.063 &  11.59(0.08) \\
  2697.346& 60 & 240 & 1.034 &   11.75(0.05)&&&  2702.367& 60 & 300 & 1.070 &  11.57(0.08) \\
  2697.350& 60 & 240 & 1.037 &   11.55(0.05)&&&  2702.371& 60 & 300 & 1.078 &  11.44(0.08) \\
  {\bf Feb. 27} &&&&&&&                          2702.375& 60 & 300 & 1.087 &  11.54(0.08) \\
  2698.252& 60 & 240 & 1.113 &   11.78(0.05)&&&  2702.380& 60 & 300 & 1.096 &  11.56(0.08) \\
  2698.256& 60 & 240 & 1.104 &   11.67(0.05)&&&  2702.386& 60 & 300 & 1.110 &  11.62(0.08) \\
  2698.260& 60 & 240 & 1.095 &   12.01(0.05)&&&  2702.391& 60 & 300 & 1.122 &  11.60(0.08) \\
  2698.264& 60 & 240 & 1.087 &   11.80(0.05)&&&  {\bf Mar. 04} &&&& \\
  2698.267& 60 & 240 & 1.080 &   11.77(0.05)&&&  2703.297& 60 & 300 & 1.033 &  11.59(0.08) \\
  2698.272& 60 & 240 & 1.073 &   11.74(0.05)&&&  2703.302& 60 & 300 & 1.030 &  11.64(0.08) \\
  2698.275& 60 & 240 & 1.067 &   11.74(0.05)&&&  2703.307& 60 & 300 & 1.029 &  11.53(0.08) \\
  {\bf Feb. 28} &&&&&&&                          2703.312& 60 & 300 & 1.029 &  11.42(0.08) \\
  2699.237& 60 & 300 & 1.149 &   11.35(0.07)&&&  2703.316& 60 & 300 & 1.029 &  11.35(0.08) \\
  2699.243& 60 & 300 & 1.132 &   11.61(0.07)&&&  2703.322& 60 & 300 & 1.030 &  11.59(0.08) \\
  2699.248& 60 & 300 & 1.118 &   11.39(0.07)&&&  2703.326& 60 & 300 & 1.032 &  11.74(0.08) \\
  2699.253& 60 & 300 & 1.106 &   11.44(0.07)&&&  2703.331& 60 & 300 & 1.035 &  11.55(0.08) \\
  2699.259& 60 & 300 & 1.092 &   11.67(0.07)&&&  2703.347& 60 & 300 & 1.048 &  11.52(0.08) \\
  2699.264& 60 & 300 & 1.083 &   11.47(0.07)&&&  2703.353& 60 & 300 & 1.055 &  11.49(0.08) \\
  2699.298& 60 & 300 & 1.038 &   11.65(0.04)&&&  2703.358& 60 & 300 & 1.062 &  11.52(0.08) \\
  2699.303& 60 & 300 & 1.035 &   11.55(0.04)&&&  2703.363& 60 & 300 & 1.069 &  11.29(0.08) \\
  2699.308& 60 & 300 & 1.033 &   11.66(0.04)&&&  2703.367& 60 & 300 & 1.077 &  11.25(0.08) \\
  2699.313& 60 & 300 & 1.031 &   11.51(0.04)&&&  2703.372& 60 & 300 & 1.085 &  11.65(0.08) \\
  2699.322& 60 & 300 & 1.029 &   11.45(0.04)&&&  2703.376& 60 & 300 & 1.095 &  11.73(0.08) \\
  2699.326& 60 & 300 & 1.029 &   11.39(0.04)&&&  2703.382& 60 & 300 & 1.107 &  11.49(0.08) \\
  2699.332& 60 & 300 & 1.030 &   11.69(0.04)&&&  {\bf Mar. 05} &&&& \\
  {\bf Mar. 02} &&&&&&&                          2704.385& 60 & 300 & 1.124 &  11.99(0.04) \\
  2701.291& 60 & 300 & 1.039 &   11.38(0.06)&&&  2704.392& 60 & 300 & 1.140 &  11.72(0.04) \\
  2701.297& 60 & 300 & 1.036 &   11.29(0.06)&&&  2704.397& 60 & 300 & 1.154 &  11.75(0.04) \\
  2701.301& 60 & 300 & 1.033 &   11.55(0.06)&&&  2704.401& 60 & 300 & 1.171 &  11.74(0.04) \\
  2701.306& 60 & 300 & 1.030 &   11.86(0.06)&&&  2704.424& 60 & 300 & 1.264 &  11.72(0.05) \\
  2701.311& 60 & 300 & 1.030 &   11.62(0.06)&&&  2704.429& 60 & 300 & 1.287 &  11.76(0.05) \\
  2701.317& 60 & 300 & 1.029 &   11.38(0.06)&&&  2704.434& 60 & 300 & 1.313 &  11.75(0.05) \\
  2701.322& 60 & 300 & 1.029 &   11.68(0.06)&&&  2704.440& 60 & 300 & 1.343 &  11.68(0.05) \\
  2701.326& 60 & 300 & 1.029 &   11.63(0.06)&&&  2704.444& 60 & 300 & 1.371 &  11.56(0.05) \\
  2701.344& 60 & 300 & 1.040 &   11.56(0.07)&&&  2704.449& 60 & 300 & 1.402 &  11.60(0.05) \\
  2701.349& 60 & 300 & 1.044 &   11.61(0.07)&&&  2704.453& 60 & 300 & 1.436 &  11.71(0.05) \\
  2701.354& 60 & 300 & 1.049 &   11.48(0.07)&&&  2704.459& 60 & 300 & 1.480 &  11.65(0.05) \\
  2701.360& 60 & 300 & 1.057 &   11.73(0.07)&&&  2704.465& 60 & 300 & 1.533 &  11.68(0.05) \\
  2701.365& 60 & 300 & 1.064 &   11.88(0.07)&&&  2704.470& 60 & 300 & 1.577 &  11.60(0.05) \\ 
  2701.370& 60 & 300 & 1.073 &   11.47(0.07)&&&  &&&& \\	
\hline \\ 
\end{tabular} 
\end{table*}

\begin{table}[h]
\caption[]{The log of observations for HD 95884 and its derived magnitudes. 
Some of the abbreviated titles are:  Exposure time is represented by Exp. time,
 Total integration time by Int. time, Air mass by A.M and $J$ band magnitude by J Mag., JD is measured 
from 2450000+}
\begin{center}
\begin{tabular}{cccccc}
\hline \hline \\
JD  &  Exp.    & Int. & A.M & Seeing & J Mag. \\
 & Time(s)&time(s)& &($''$)& (Error)   \\\hline
2696.410 & 0.4 & 3.6 & 1.049 & 1.7 & 6.75(0.02) \\
2697.304 & 2.0 & 8.0 & 1.039 & 1.6 & 6.67(0.05) \\
2698.278 & 2.0 & 8.0 & 1.060 & 1.6 & 6.67(0.04) \\
2699.229 & 2.0 & 18.0 & 1.172 & 1.8 &6.67(0.06) \\
2699.277 & 2.0 & 18.0 & 1.061 & 1.6  &6.69(0.05) \\
2699.290 & 2.0 & 18.0 & 1.045 & 1.7 &6.71(0.06) \\
2701.288 & 2.0 & 18.0 & 1.044 & 1.9 &6.62(0.12) \\
2701.338 & 2.0 & 18.0 & 1.040 & 1.7 &6.71(0.16) \\
2702.297 & 2.0 & 18.0 & 1.035 & 1.9 &6.76(0.09) \\
2702.351 & 2.0 & 18.0 & 1.055 & 1.8 &6.75(0.07) \\
2703.272 & 2.0 & 18.0 & 1.053 & 1.5 &6.64(0.15) \\
2704.377 & 2.0 & 10.0 & 1.111 & 1.3 &6.77(0.02) \\
2704.419 & 2.0 & 10.0 & 1.253 & 1.5 &6.75(0.01) \\
\hline 
\end{tabular}
\end{center} 
\end{table}

Photometry in the $J$ band was  done at the  Mt. Abu 1.2m  telescope using a Near Infrared 
Imager/Spectrometer with a 256$\times$256 HgCdTe NICMOS 3  array. The instrument was used in the 
imaging mode with a $\sim$ 2 $\times$ 2 arcmin$^{2}$ field. Mrk 421 was observed continuously on 
all nights between 25 February and 5 March 2003 except on March 01, 2003 when unfavorable sky conditions 
did not permit observations. The sky was  photometric during all epochs of observations. It
would have been preferable if the object, standard and comparison stars were all present in the
same image frame - this would have minimised the errors caused by sky transparency and seeing
on the derived magnitudes of Mrk 421. Since there are no bright stars in the $\sim$ 2 $\times$ 2 
arcmin$^{2}$ Mrk 421 field that can serve this purpose, we have chosen very nearby standard 
and comparison stars to circumvent this problem. The UKIRT standard star HD 105601 ($J$=6.821) was used 
for photometric calibration  for all  observations. Also, as a reliability check for 
any observed  variations in Mrk 421, the comparison star HD 95884 was  observed.  Mrk421, HD 105601 
and HD 95884 were all observed at similar airmass to minimize atmospheric extinction 
corrections in their derived magnitudes. Observations of the comparison and standard stars were 
done just before or immediately after the observations of Mrk 421. The standard and comparison stars  
were observed once
during the course of the night on four epochs viz. 25, 26, 27 February and 4 March, twice 
during 2, 3 and 5 March and thrice on 28 February.  The sequence of  observations, 
for each of the standard/comparison stars   and Mrk 421, involved the following procedure. Several 
images were obtained, in at least four  dithered positions, offset by approximately 
30 arcseconds. The dithered frames were median-combined (using IRAF) to generate the corresponding 
sky frame which was then subtracted from the object frames. Aperture photometry was done at each 
dithered position  using the APPHOT task from IRAF to yield an instrumental magnitude. Since the 
observation cycle is small for  the standard and comparison stars (both being bright), their derived 
magnitudes  from the different dithered positions were averaged to yield a mean value. The aperture 
size, during aperture photometry,  was chosen to be generally four times the FWHM of the recorded 
stellar image (i.e typically 6-8 arcseconds). The   atmospheric  correction for the instrumental 
magnitudes was done using an average extinction coefficient of k$_{j}$  = 0.15 for the $J$ band for 
the Mt. Abu  Observatory site.  A detailed log of the observations and the derived magnitudes for 
Mrk 421  is presented in  Table 1. Similar details for the comparison star HD 95884 are given in 
Table 2. From  Table 2 it is seen that the mean  $J$ magnitude of HD 95884 of 6.705 agrees well with 
its 2MASS magnitude of 6.693.

\section{Results and Discussion}

We have shown in Fig. 1 the $J$ band light curve of Mrk 421 on the individual nights of 
observation. The periods of overlap of our data with RXTE observations are also indicated. From 
our data it is seen that  the  maximum variation of the source is 0.89 magnitudes (between its 
brightest level at 11.12 mag on  JD 2452696.399 and faintest level at 12.01 mag on JD 2452698.260). 
In comparison, the data from Takalo et al. (1992) show a maximum change of 0.32 mags during their 
observations while the Makino et al. (1987) observations show a marginal change of only 0.03 mags. 
This shows that the source was quite active during our epoch of observations. In terms of flux 
variation, the observed 0.89 magnitude change corresponds to a large 84 percent peak-to-peak 
variation in the average $J$ band brightness of Mrk 421 (the average $J$ magnitude for the 95 
observed data points is 11.559).  It  also appears that the faintest level of the source detected 
by us (12.01 magnitude) could might be possibly be  the faintest recorded magnitude for this object (by 
comparing with data from Fan \& Lin; 1999). The maximum brightness variation recorded on the 
individual nights (in sequence of observation date) are  0.15, 0.29, 0.34, 0.34, 0.59, 0.47, 0.49 
and 0.39 magnitudes respectively. Thus considerable IDV in Mrk 421 is seen on all nights. As 
a comparison we note that the observations of Kidger et al. (1999) over a 3 hour period show a 
maximum variability of $\sim$ 0.1 magnitude.

\begin{figure}[t]
\includegraphics[bb=18 144 592 718,width=4.3in,height=5.6in,clip]{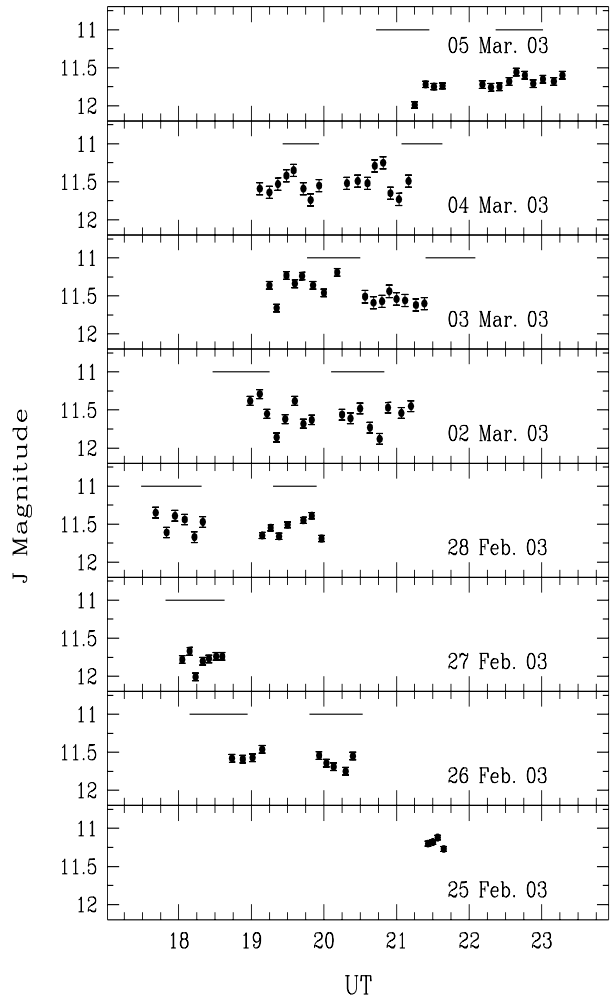}
\caption[]{ The $J$ band light curve of Mrk 421 during Feb. 25 - March 05,
2003. Data for an individual night are plotted in one panel. We have also indicated, by horizontal 
lines, the periods for which there is a temporal overlap between the present data and 
the RXTE observation slots.} 
\label{fig1}
\end{figure}

\begin{figure}[h]
\includegraphics[bb=18 144 592 718,width=3.5in,height=3.5in,clip]{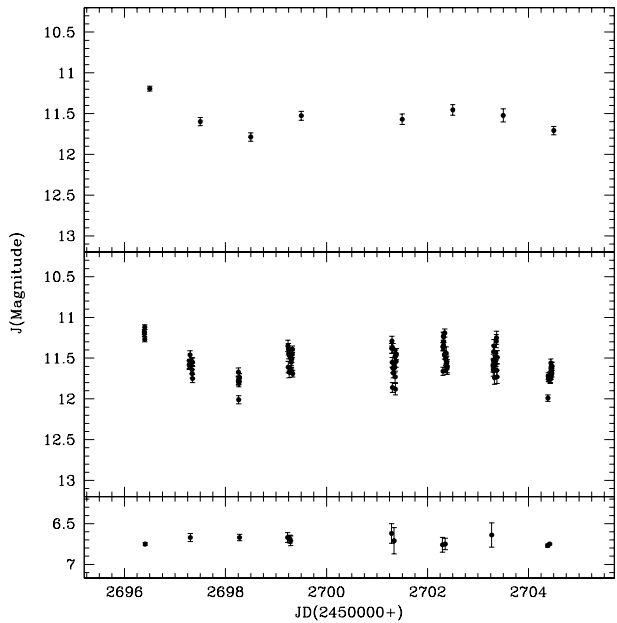}
\caption[]{ The upper panel shows the mean $J$ magnitude on 8 nights (between 25 Feb. to 5 March 
2003) for Mrk 421. The  middle panel of the above plot shows the complete set of data points of our 
observations. The lowest panel shows the mean $J$ magnitudes for the comparison star HD 95884.}
\label{fig2}
\end{figure}

The short term variability is best seen from Fig. 2 where we have plotted average brightness 
values on the individual nights in the upper panel. The middle panel shows all the data of Fig. 1 
in a more compact and comprehensive format to enable comparison with the upper panel. The lowest panel 
of Fig. 2, which shows the near-constancy of the observed magnitudes for the comparison star HD 95884, 
can be used to judge the validity of the observed variability in Mrk 421. Figure 2 
(top panel)  shows that there are  short term variations during 25-28 February, a near constant 
phase between 2-4 March followed by a slight decline on 5 March. Some amount of flaring activity 
is seen in Mrk 421. On 2 March a flare is seen at JD 2452701.317, with a brightness excursion of 
0.48 magnitudes. Similarly a flare is seen on 3 March and two flares on 4 March at 
JD 2452702.341, JD 2452703.316 and JD 2452703.367  with variations of 0.45, 0.29 and 0.49 magnitudes 
respectively. Further, the data of Feb. 25 suggest that a flare could have just preceded our first 
data point on that date since the source was seen to be brightening and reached its brightest level 
on that date. It may be mentioned that Takalo et al. (1992) have also detected two instances of 
flares in Mrk 421 with $J$ band variations  of 0.25 and 0.2 magnitudes.

It is useful to examine how the observed $J$ magnitudes of Mrk 421
are affected by variations in the seeing conditions. A variation in
seeing can lead to a change in the observed brightness of Mrk 421
because of  differing brightness contributions - within a measuring
aperture diameter - from the host galaxy that surrounds Mrk 421. A
detailed study in this context has been done by Nilsson et al. (1999)
in the $R$ band for three BL Lac objects including Mrk 421. While their
$R$ band results cannot be extrapolated directly to the $J$ band
observations reported here, it is still instructive to make an 
assessment of the effects of seeing on the  $J$ magnitudes reported here.
From the compiled data of Nilsson et al. (1999; refer Table 6), by
choosing  an aperture diameter of 7.5$\arcsec$ - which is quite represenative
of the aperture used in the present studies - it is seen that a
variation in the seeing from 2$\arcsec$ to 4$\arcsec$ changes the relative contribution
of the host galaxy to the observed $R$ magnitude of Mrk 421 from 0.04 to
0.18 magnitudes i.e. a change of 0.14 magnitudes for a 2$\arcsec$ change in
seeing. In this respect we  have given the seeing value in arc seconds
for the comparison star HD 95884 in Table 2. The seeing was estimated
by measuring the full width at half maximum of the stellar images of
this star. The seeing values during the Mrk 421 observations can be
assumed to be similar to those of HD 95884 since observations of both
objects were separated by  small time differences and were also made
at similar airmass. As may be seen from Table 2, the seeing values
were fairly constant and did not vary significantly during the different
epochs of Mrk 421 observation. Thus we would infer that the small,
observed changes in the seeing could have only a marginal effect on the
observed $J$ magnitudes. The above arguments would therefore indicate
that the  observed short-term and intra-day variability in Mrk 421 - the
central point of this study - that are  reported here are reliable. It is
however outside the scope of this work to construct two-dimensional
photometric maps of Mrk 421 to ``clean" the observed $J$ band  magnitudes
of the host galaxy contribution as has been done in the detailed analysis
by Nilsson et al. (1999).

From different multiwavelength campaigns it has been seen that strong optical variations are 
generally not seen during X-ray and $\gamma$-ray flaring events (Tosti et al. 1998 and references 
therein). However, there is some evidence that optical flares are generally accompanied by X-ray and 
$\gamma$-ray flares (Hartman et al. 2001). A similar correlated variability is also seen for the 
radio region.  Katarzy$\acute n$ski et al. (2003) shows a well defined correlation between observed radio 
outbursts in Mrk 421 with a corresponding X-ray outburst and a $\gamma$-ray flare in the TeV range. 
Since  considerable flaring activity is seen in the present $J$ band data, it will be interesting 
to see whether there is correlated variability in the X-ray/$\gamma$-ray regimes.  RXTE and STACEE 
observations, which have been done simultaneously, can confirm this.

There are several models that explain the intraday and short term variability in blazars viz. 
shock-in-jet models, accretion-disk based models and plasma-process related models that can explain 
IDV over an extended range of wavelengths (Wagner \& Witzel 1995, Urry \& Padovani 1995 and references 
therein). The observed variability in our present data - taken by itself - could be consistently 
explained by any successful model for IDV. However, when taken in conjunction with X-ray/$\gamma$-ray 
data  it may provide a more definite insight into the observed IDV in Mrk 421.
 
\begin{acknowledgements}

We thank the referee  L. O. Takalo for constructive comments that have helped us to improve
the paper. We are also grateful to P. N. Bhat, Tata Institute of Fundamental Research, Mumbai for
bringing the RXTE observational campaign to our notice. The research work at Physical Research 
Laboratory is funded by the Department of Space, Government of India and at Harish-Chandra Research 
Institute  by the Department of Atomic Energy, Government of India. IRAF is distributed by NOAO, USA. 

\end{acknowledgements}

\end{document}